# BERING – A DEEP SPACE MISSION TO STUDY THE ORIGIN AND EVOLUTION OF THE ASTEROID MAIN BELT

**Philip R. Bidstrup[1], Henning Haack[1], Anja C. Andersen[2,3], Rene Michelsen[4], and John Leif Jørgensen[4]**

[1]Geological Museum, University of Copenhagen, Øster Voldgade 5-7, 1350 Copenhagen K, Denmark, philipb@snm.ku.dk, hh@snm.ku.dk Ph +45 35322345, Fax: +45 35322325.
[2]Dark Cosmology Center, Niels Bohr Institute, University of Copenhagen, Juliane Maries Vej 30, 2100 Copenhagen, Denmark, anja@astro.ku.dk  [3]NORDITA, Blegdamsvej 17, 2100 Copenhagen Ø, Denmark
[4]Ørsted, Danish Technical University, Bldg 327, 2800 Lyngby, Denmark, rm@oersted.dtu.dk, jlj@oersted.dtu.dk

## ABSTRACT

Using a fully autonomous spacecraft - Bering - we propose to detect and study sub-km asteroids from an orbit within the asteroid Main Belt. The main purpose of the proposed Bering mission is to detect a statistically significant sample of an expected population of approximately $10^{10}$ main belt asteroids in the size range 1 m to 1 km. These asteroids are too faint to be observed using Earth-based telescopes. Sub-km asteroids can be detected from spacecraft at close range but due to the high relative velocities and the long communication times this requires a fully autonomous spacecraft. Using theoretical estimates of the distribution and abundance of sub-km asteroids we find that the Bering mission would detect approximately 6 new sub-km asteroids per day. With an expected lifetime for the mission of a few years we expect to detect and study several thousand sub-km asteroids. Results from the Bering mission would allow us to: 1) Provide further links between groups of meteorites and their parent asteroids. 2) Constrain the cratering rate at planetary surfaces and thus allow significantly improved cratering ages for terrains on Mars and other planets. 3) Constrain processes that transfer small asteroids from orbits in the main belt to the inner Solar System.

## 1. INTRODUCTION

The discovery of Ceres in 1801 was the first of many asteroid discoveries and ongoing systematic searches for asteroids have, to date, brought the number of observed asteroids to almost 250.000 asteroids [9]. The asteroids have been observed through many size-ranges with an increasing number of smaller asteroids, yet it is estimated that approximately $10^{10}$ Main Belt asteroids in the size-range of 1 m to 1 km should exist [3]. Due to the smallest asteroids faint appearance, as seen from the Earth, Earth-based telescopes have so far been unable to observe the predicted small asteroid population in the asteroid Main Belt. However, discoveries of small near Earth asteroids (NEAs) hint the existence of similar small Main Belt asteroids as NEAs are thought to origin from the asteroid Main Belt.

Most of the smallest asteroids are fragments from close encounters or impact events with larger asteroids in the Main Belt [3]. Despite their small size, the fragments are of considerable interest. As fragments, the small asteroids have younger surfaces than larger asteroids and recently produced fragments have only a minimal regolith cover due to a minimal exposure to space weathering effects. Spectral signatures of these recent fragments can be compared to the spectral signatures of the sample of meteorites collected on the surface of the Earth and can make it possible to match meteorites with asteroid fragments. Following the trail of fragments back to their parent asteroid will ultimately allow us use the meteoritic evidence to understand the distribution of asteroid types in the Main Belt and thus shed light on the early evolution of the Solar System. In addition, in order to constrain crater count ages of planetary surfaces, including ages inferred for the recent climatic evolution of Mars, observations of small asteroids are needed. The vast majority of recent craters on Mars are due to impacts of sub-km asteroids and age estimates [5] therefore rely on poorly constrained estimates of the abundance of small objects. Finally, estimating the actual size distribution of small asteroids will allow us to constrain models of the collisional evolution of Main Belt asteroids [3].

## 2. THE BERING MISSION

### 2.1 Concept

We propose to detect and study a statistically significant sample of sub-km asteroids using a fully autonomous spacecraft observing from within the asteroid Main Belt.

The spacecraft – named Bering after the Danish explorer Vitus Bering – is designed to be of octagonal cylindrical shape of 2 m in diameter and 1.75 m in height. It will mainly consist of the primary science instruments: an array of 7 micro Advanced Stellar Compasses (μASC star trackers), a folding mirror based science telescope and a multi-spectral imager. Instruments can be seen mounted on a concept of the spacecraft structure from Fig. 1. The Bering spacecraft will be detecting and observing asteroids once it reaches its designated orbit inside the asteroid Main Belt.

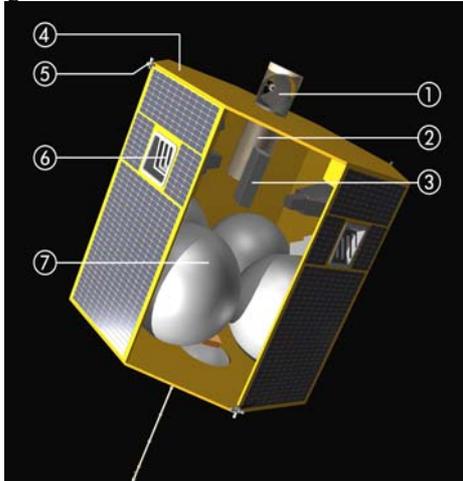

Fig. 1. The Bering spacecraft structure concept: 1) Folding Mirror 2) Telescope 3) 100K box for a infrared part of the multi-spectral imager 4) S/C top 5) ACS Thrusters 6) µASC Camera 7) Propulsion tanks [10].

However, also during transfer of the spacecraft with a Venus flyby through the inner solar system, NEAs can be observed. The transfer and final orbits can be seen in Fig. 2. Different from any other mission to the asteroids, Bering is a survey mission. The goals are not to gather highly specific knowledge of a few asteroids, but to gain broad information of as many asteroids in the Main Belt as possible. Furthermore, its purposes seek to fulfil exploration and discovery from a yet unknown population of sub-kilometre sized asteroids.

**2.2 The Bering science objectives:**

Primary science objectives:
- Sampling of the NEA and Main Belt asteroid size-distribution by discovery of new sub-kilometre asteroids.
- Follow-up and orbit procurement of new discoveries.
- A rough taxonomic classification of the asteroids by multicolour photometry.

Secondary science objectives based on long-term observations are restricted to campaign modes, and may not be applicable to all discoveries:
- Follow-up at faint magnitudes of new discoveries.
- Photometric light curves, to provide information on the spin of the asteroids.
- Photometry over several phase angles, to provide information on shape and albedo.
- Detection of satellites around known asteroids.
- Determination of mass/density of binary asteroids.

Besides the goals for statistical sampling, fly-by of a few selected targets could drastically augment the scientific return in terms of measurement of mass-, magnetic- and surface properties.

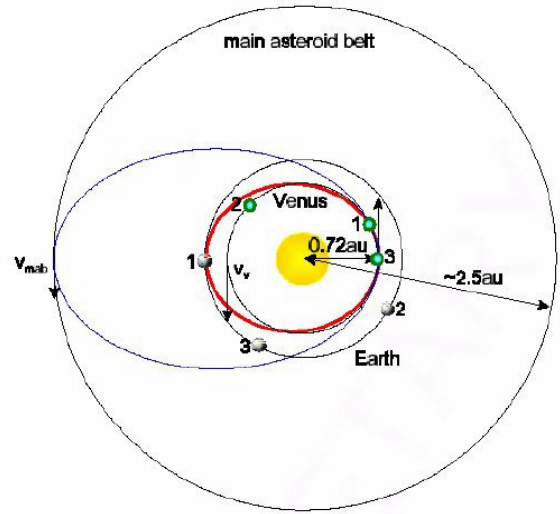

Fig. 2. Transfer orbit and designated orbit of the Bering spacecraft. Two final orbits are in consideration: a circular orbit inside the asteroid Main Belt at 2.5 AU from the Sun and the uncircularised elliptic orbit with aphelion at 2.5 AU [1].

**2.3 The micro Advanced Stellar Compass**

The detection of asteroids is based on six of the seven mounted micro Advanced Stellar Compasses (µASCs), being the backbone technology of the Bering mission. The µASC was initially constructed by Ørsted•DTU for determination of the orientation of satellites by position matching of recorded stars from pictures of the stellar background with onboard star catalogues [6]. Asteroids and other objects not accounted for in the star catalogue, will appear on the recordings and can be filtered out. By re-observation of the same areas of the sky uncatalogued objects that do not show motion can be removed through a second filtering and the remaining objects will be down linked to Earth for identification and further calculations. Everything to this point is done autonomously. The strength of the µASC technology is exactly that it is fully autonomous, and will detect asteroids without ground based control. Because the closed loop communication to the Bering satellite is more than 30 minutes, an autonomous system is absolutely necessary as the asteroids of interest are small and faint and will be detected at much closer ranges than larger asteroids, thus moving through the sphere of detection more rapidly.

Four µASC camera head units (CHUs) will be mounted on four of the eight sides of the spacecraft, two CHUs will be placed on the top and one CHU will be used to aid the science telescope to ensure a higher accuracy.

By combining the several CHUs on the slowly spinning spacecraft (around its axis of symmetry), an almost full coverage of the celestial sphere is obtained [7].

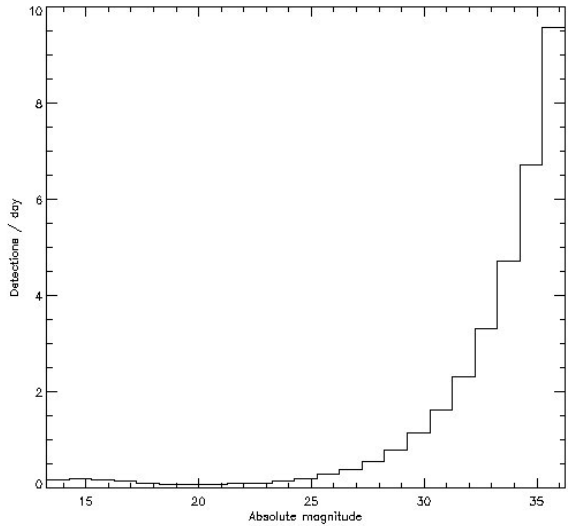

Fig. 3. The estimated detection rates per day (without trailing losses) against the absolute magnitude of the detected asteroids. The range of the absolute magnitude scale corresponds to the diameters ranging from of kilometre-sized asteroids (small absolute magnitudes) to asteroids 10 cm in diameter (large absolute magnitudes). The spacecraft was assumed to be in a circular orbit at 2.5 AU with detection limit of $m_v=13$ during the simulation.

The μASC is highly robust and radiation-proof, and has extensive flight heritage with for instance the missions: Ørsted, Champ, Proba, Grace, Adeos-2 and Smart-1.

## 3 DETECTION OF ASTEROIDS

### 3.1 Unknown asteroids

The Bering spacecraft will be observing asteroids that are unknown from an unknown asteroid population which makes the configuration of the instruments a difficult task. Most important, it is necessary to understand the basis of the detections and for this an asteroid detection model has been made to estimate the detection rates for different sizes of asteroids. The detection rate is controlled by two factors – the limiting magnitude and the tangential velocity of the objects.

### 3.2 Asteroid detection model

Assuming that the velocity distribution of the small unknown asteroids is the same as for the larger known asteroids, the velocity distribution relative to the Bering spacecraft in a circular orbit 2.5 AU from the Sun can be created from the asteroid data available at the minor planet center (MPC) [9]. Calculating the density of known asteroids as function of heliocentric distance and height above the ecliptic, we can extend the densities to include the small asteroids predicted from models of the size-distribution. This is done assuming the small asteroids are distributed similar to the larger known asteroids. The number of asteroids entering the detection range based on the chosen detection limit (e.g. $m_v=13$) during some time-interval, can be calculated as function of asteroid size. The model does not yet include the phase of the asteroids and assumes phase angles of 0°. The size-specific detection rates have been calculated for $m_v=13$ and can be seen from Fig. 3. It is worth noticing that even though the small asteroids can only be detected at small distances; they are, as expected, abundant enough to dominate the larger fraction of the detection rates. A numerical approach to the detection rates of the Bering spacecraft in a circular orbit has previously been made and gives a good correspondence with the results of the statistical model [8].

### 3.3 Trailing losses

While large asteroids can be detected from large distances, their angular velocities will be observed as small because the angular velocity is proportional to the inverse of distance. Due to the small asteroids small amount of reflected sunlight, they will only be detected at close range and will thus have high angular velocities. Some of the asteroids will have sufficiently high angular velocities that they escape detection due to smearing of their recorded image. This is a known problem from ground based observations, where the sensitivity of the detecting equipment grows with the time of image integration. However, as the sensitivity of the μASC can be configured to a detection limit of $m_v=13$, it will have integration times of only a couple of seconds and can thus record asteroids with very high angular velocities.

It can be seen from Table 1 how a majority of the detections are lost due to trailing losses, but also that it is from detection with a high sensitivity that the trailing losses are most dominant. Lower sensitivity demands only small integration times and trailing losses will thus play a smaller role. The current technical limit of the μASC detection system as modified for the Bering mission is $m_v=13$, however, the configuration can be altered in-flight if a special campaign mode e.g. should require minimal trailing losses.

Table 1. The calculated detection rates for different choices of detection limit. Last column is corrected for trailing losses. The current technical limit is $m_v \approx 13$.

| $m_v$ | Detections/day uncorrected | Detections/day trailing loss corrected |
|---|---|---|
| 14 | 82.6 | 10.7 |

| | | | | | |
|---|---|---|---|---|---|
| 13 | 32.9 | 5.7 | 10 | 2.1 | 0.9 |
| 12 | 13.1 | 3.1 | | | |
| 11 | 5.2 | 1.7 | | | |

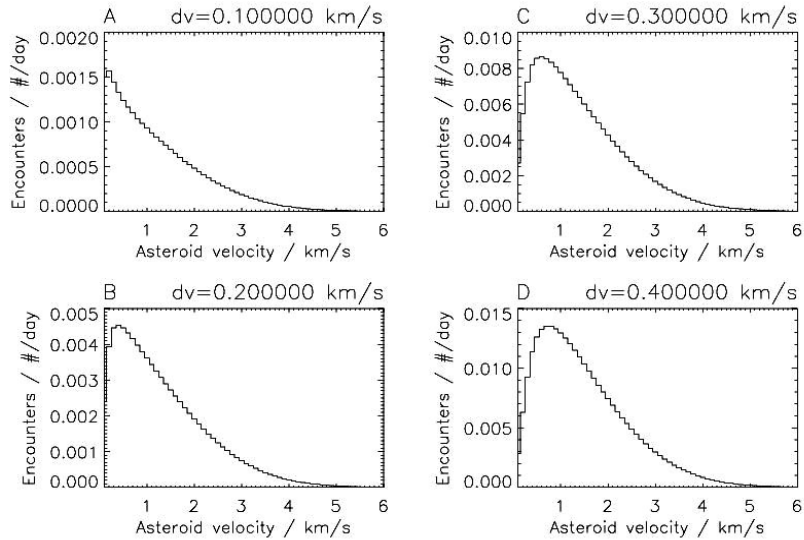

Fig. 4. Calculations for four different velocity changes (Δv). The first axis shows the incoming asteroids' relative velocities and the second axis shows the number of asteroids that could be closely encountered per day with the given boost. It seems that asteroids with a relative velocity of ~0.5 km/s are the most frequent targets for close encounters for velocity changes of the given sizes. As the Δv-capability grows from panel A to D, asteroids with faster relative velocities can be reached. The total flux of possible close encounters is the integral of the graph and the time until a possible encounter is the inverse of the flux. Detection limit used for calculations is $m_v$=13.

## 4 ASTEROID OBSERVATIONS

### 4.1 The Science telescope

Once an asteroid is detected and an estimated orbit has been determined from the μASC data, the science telescope will continue the observation. With a seventh μASC mounted on the secondary mirror of the science telescope, it will accurately guide the telescope to its target [4]. Guided, the science telescope can follow asteroids for a long period of time and thereby optimise their orbit estimates. Also, during this time, a rough taxonomic classification of the asteroids can be made with the spectral imager. Furthermore, repeated follow-up of selected targets can fulfil part of the secondary goals of the mission such as light-curves and a study of phase angles can reveal spin, shape and albedo of the asteroids.

While some objectives can be obtained from the suggested spacecraft orbital path, some can not due to the fact that asteroids will almost always pass by at large distances. Only a couple of asteroids will, during mission lifetime of approximately 3-5 years, pass so close-by that images of high resolution are recorded [2]. Most asteroids will be recorded extending less than 10x10 CCD pixels. With only a few asteroids being recorded in high resolution, it is very unlikely that a close encounter with an asteroid will occur by chance.

### 4.2 Fly-by of asteroids

While it is improbable that close encounters should occur by chance, the orbital path need not much change in order to organise the encounter.

An onboard cold gas propulsion system is used for the spacecraft attitude control, but it can also provide a spacecraft velocity change (Δv). Applied correctly, this Δv can ensure a close encounter that will not only enhance the quality of observations but also enable several additional science goals, e.g. the measurement of magnetic properties, laser-ranging, search for asteroid satellites and mass determination of binary asteroids etc. Once the orbit of a newly discovered asteroid has been determined, thrusters can deliver the needed Δv towards the desired point of rendezvous. Depending on configuration, all allocated propulsion can be used for a single target or the thrust can be divided into several smaller boosts. More small boosts will give the opportunity of several flybys, but as the Δv delivered will be smaller not as many asteroids will fulfil the requirements for the close approach. Calculations with a detection limit of $m_V$=13 [2] show that for the large boost of Δv=1.0 km/s it will take approximately 2-3 days before a close approach opportunity is present and for Δv=0.01 km/s it will take on the order of 100 days. See Fig. 4 for details. In this way, depending on the time available, several flybys can be accomplished if one is willing to wait for the opportunity.

## 5 CONCLUSION

The Bering mission proposes to detect and study a significant sample of the estimated $10^{10}$ unknown sub-km Main Belt asteroids from within the asteroid Main Belt. Using micro Advanced Stellar Compasses (µASCs) for autonomous asteroid detection, the Bering spacecraft can with a present technological detection limit of $m_v$=13 detect ~6 new asteroids per day from a circular orbit at 2.5 AU from the Sun. With an estimated lifetime of the spacecraft of several years we should expect a total of 5-10.000 new small objects.

The study of asteroids will consist of the detection of the new small asteroids which will contribute to the size-distribution that presently is incomplete with respect to sub-km asteroids. Furthermore, the studies will contain a determination of the discovered asteroids orbital parameters so that follow-up science can later be applied from space or by ground based observations. With the multi-colour imager a rough taxonomic classification can show whether the small asteroids are distributed with same differentiation in asteroid compositional types as known for the larger asteroids. Perhaps the spectral signature can also reveal a match between recovered meteorites and the small asteroids composition – a match that is unsuccessful with the larger asteroids due to space weathering effects. These effects are believed to be minimal with the small asteroids.

During spacecraft campaign modes, follow-up observation of selected asteroids can deliver light-curves and photometry over several phase angles, and thus reveal information such as asteroid spin, shape and albedo. The mission can be extended to address fly-bys of one or more asteroids which will enhance the outcome with additional information of mass, magnetics and high resolution images.

The Bering mission is unique in its method of discovering the smaller of the asteroids in the Main Belt that can only be discovered from space since the asteroids are too faint and fast to be detected from Earth. Compared to other asteroid search programs, Bering will detect Main Belt asteroids that are smaller than ever detected before, see Fig. 5 for comparison.

## ACKNOWLEDGEMENT

Dark Cosmology Center is funded by the Danish National Research Foundation.

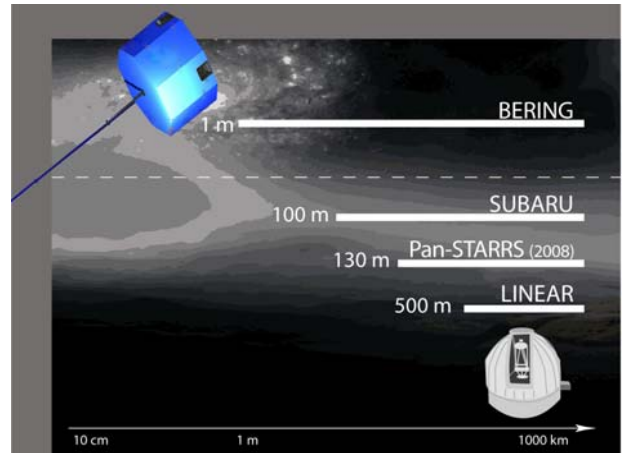

Fig. 5. The size range of detectable Main Belt asteroids with Bering and with a number of high-performing ground based telescopes. LINEAR is a currently existing asteroid survey using a 1m telescope. The Pan-STARRS project (full initiation expected in 2008) is a dedicated asteroid survey, and will consist of four 1.8m telescopes. Subaru is an 8.2m astronomical telescope and is among the most powerful ground based telescopes, although it is not dedicated to surveying asteroids